# Economic complexity and capabilities of Indian States

Anand Sahasranaman[1,2,*] and Henrik Jeldtoft Jensen[2,3,#]

[1]Division of Science and Division of Social Science, Krea University, Sri City, AP 517646, India.

[2]Centre for Complexity Science and Dept of Mathematics, Imperial College London, London SW72AZ, UK.

[3]Institute of Innovative Research, Tokyo Institute of Technology, 4259, Nagatsuta-cho, Yokohama 226-8502, Japan.

[*] Corresponding Author. Email: anand.sahasranaman@krea.edu.in

[#] Email: h.jensen@imperial.ac.uk

**Abstract:**

We explore economic competitiveness of Indian states based on the economic complexity algorithm, using a pair of coupled non-linear maps to characterize the Fitness of states and Complexity of products exported by them. We find that states produce almost all products within their productive capabilities - diversifying rather than specializing, and that the probability of coexistence of any pair of productive capabilities is maximized when capabilities are of similar complexity. Therefore, states require long time horizons to build complex capabilities and diverse products. We contextualize Fitness using human development, and find an emergent typology of states. Of most concern are the states of Odisha, Uttar Pradesh, and Bihar, stuck in vicious feedback cycles of poor economic complexity and low human development. Economic complexity also reveals significant concerns with the economic trajectories of Punjab, Gujarat, and West Bengal. We discuss these emergent trends within the framework of India's modern economic history.

# 1. Introduction:

There are a multitude of economic theories attempting to explain the evolution of economic development paths of nations (Smith, 1776; Ricardo, 1817; Solow, 1956; Swan, 1956; Romer, 1986; Lucas, 1988; Aghion & Howitt, 1992). Classical economic theory focused on national income as the basis for economic development and considered critical three factors of production – labour, land and capital (Smith, 1776). It also stressed the importance of trade to economic growth, with Ricardo outlining the principle of comparative advantage, according to which nations exported only those goods and services they could produce at lower opportunity cost, ultimately resulting in specialization of national economic production systems (Ricardo, 1817). Modern neoclassical theory, typified by the Solow-Swan model (Solow, 1956; Swan, 1956), emphasised the importance of technological change in addition to labour and capital, and predicted long run convergence to a steady-state equilibrium. A natural outcome of this theory is that permanent growth remains possible only through continuous technological innovation. While the neoclassical approach remains influential, dissatisfaction with its assumption of technology as an exogenous effect led to the development of endogenous growth theory (Romer, 1986; Lucas, 1988; Aghion & Howitt, 1992), which endogenized productivity within the growth model. This has been accomplished, for instance, by modelling accumulation of knowledge as investment in human capital and considering its spillover effects in generating growth (Romer, 1986; Lucas, 1988), or alternatively by modelling industrial innovations to improve products - thus incorporating the factor of obsolescence - and essentially embodying Schumpeter's vision of growth (Schumpeter, 1942) through creative destruction (Aghion & Howitt, 1992).

It has been argued that the predictive power of these theoretical models is compromised on account of simplifying assumptions such as equilibrating outcomes and rational agents, that prevent a dynamical representation of the economic system (Miller & Page, 2007). It was

Hayek who made the case that the generation of market outcomes as a consequence of interactions between economic agents was essential to understanding the truly significant aspects of the emergent market behaviour (Hayek, 1945). There has indeed been a long realization of the need for dynamic representation of social phenomena (von Neumann & Morgenstern, 2007), and even predictions that the widespread use of computers could increasingly influence the development of theory behind such complex systems (Weaver, 1991; Ulam, 1991). Essentially, processes such as economic growth are well suited to exploration as emergent phenomena arising out of complex and heterogenous interactions across multiple scales.

In this context, the recent emergence of Economic Complexity as the basis to explore the economic capabilities and growth prospects of nations is significant (Hidalgo & Hausmann, 2009; Hidalgo, Klinger, Barabási, & Hausmann, 2007; Tacchella, Cristelli, Caldarelli, Gabrielli, & Pietronero, 2012; Pietronero, Cristelli, & Tacchella, 2013; Cristelli, Tacchella, Cader, Roster, & Pietronero, 2017). The underlying thesis of Economic Complexity is that the productivity of a nation is a function of its underlying non-tradable 'capabilities' (such as infrastructure, regulations, and skills) and that differences in national economic performance are explained by differences in economic complexity as encapsulated by the diversity of and interactions between these capabilities (Hidalgo & Hausmann, 2009; Hidalgo, Klinger, Barabási, & Hausmann, 2007).

While economic complexity has generally been used in the context of economic performance of nations, we argue that it provides us a useful framework to explore the economic capabilities of Indian states (sub-national administrative regions). India is a nation of sub-continental scale with vast cultural, social, and economic diversity – it has 30 languages spoken by more than a million people and distinct socio-cultural practices across geographies (Census of India, 2011). Importantly, it is home to one-sixth of humanity, with a strongly federal structure where significant responsibility for socio-economic development is wielded by state governments (Ahluwalia, 2000). India's post-independence economic history has been characterized by significant heterogeneity in economic performance across states, with southern and western states exhibiting much stronger performance than the states of the Indo-Gangetic plain (Bose, 1988; Sharma, 2015; Kurian, 2000; Bardhan, 2012). Over this time, income disparities across states have only increased, with no evidence for convergence (Datt & Ravallion, 2002; Marjit & Mitra, 1996; Ghate, 2008). Given this context, we argue that it is reasonable, and indeed valuable, to consider the Indian economic system as being driven by the capabilities of its constituent states, and to consequently develop a deeper understanding of the nature of economic capabilities of these states.

Specifically, in this work, we propose to use the framework of economic complexity to explore the performance of Indian states based on their export data. We also check for the robustness of the fitness measures of states resulting from the economic complexity algorithm. In order to contextualize fitness, we propose to study the co-evolution of fitness measures and human development indicators of states over time, to further our understanding of how the process of economic development, as revealed by economic complexity, changes with human development in India. Finally, we discuss specific results for states and significant exceptions thrown up by economic complexity, in the context of India's modern economic history.

## 2. Data and Methods:

Hidalgo and Hausmann (2009) show that it is indeed possible to infer the diversity and ubiquity of capabilities of countries by merely looking at their export baskets and interpreting the product-country data as a bipartite graph, which they contend is the decomposition of a tripartite graph where countries are linked to capabilities they possess and products to the capabilities required to produce them. The economic complexity algorithm characterizes the structure of the bipartite graph through an iterative procedure to produce a symmetric set of variables for the two kinds of nodes on the network – one set to describe the ubiquity of products based on the number of countries producing them and the other to describe the diversity of countries based on the basket of products they export. The actual mechanism we use to implement the economic complexity algorithm is based on Tacchella, Cristelli, Caldarelli, Gabrielli, & Pietronero (2012), who proposed a non-linear, iterative approach, in the spirit of PageRank (Page, Brin, Motwani, & Winograd, 1999), to measure the Complexity of products and the Fitness of countries that produced them - as the fixed point of the iteration of two non-linear coupled equations. In this context, the Complexity of a product is best understood as a measure of the capabilities required to produce the product and the Fitness of a nation is a measure of its competitiveness and adaptability (or reservoir of capabilities).

The first step in the execution of the economic complexity algorithm is the construction of the state-product network as an adjacency matrix ($M$). Each term of the matrix, $M_{sp}$ - corresponding to state $s$ and product $p$, is a measure of the Revealed Comparative Advantage ($RCA_{sp}$) (Balassa, 1965) of state $s$ in product $p$. If $q_{sp}$ is the amount (in monetary terms) of the export by state $s$ of product $p$, then $RCA_{sp}$ is defined as (Eq. 1):

$$RCA_{sp} = \frac{\frac{q_{sp}}{\Sigma_i q_{s_i p}}}{\frac{\Sigma_j q_{sp_j}}{\Sigma_i \Sigma_j q_{s_i p_j}}} \qquad (1)$$

$RCA_{sp}$ is meant to be a measure of the relative advantage that state $s$ has in producing product $p$. Given $RCA_{sp}$, the corresponding entry ($M_{sp}$) in the adjacency matrix is (Eq. 2):

$$M_{sp} = \begin{cases} 1, & if\ RCA_{sp} \geq 1 \\ 0, & if\ RCA_{sp} < 1 \end{cases} \qquad (2)$$

Once we have used this mechanism to construct adjacency matrix ($M$), we defined the iteration process which couples the Fitness of a state ($F_s$) with the Complexity of a product ($Q_p$) as follows:

$$F_s^n = \frac{\sum_p M_{sp} Q_p^{n-1}}{\langle \sum_p M_{sp} Q_p^{n-1} \rangle_s} \qquad (3)$$

$$Q_p^n = \frac{\sum_s \frac{1}{\frac{M_{sp}}{F_s^{n-1}}}}{\langle \sum_s \frac{1}{\frac{M_{sp}}{F_s^{n-1}}} \rangle_p} \qquad (4)$$

At each iteration ($n$), Fitness of a state ($F_s$) is proportional to the linear sum of the complexity of products in its export basket (Eq. 3), while Complexity of a product ($Q_p$) weights the Fitness of producer states in a non-linear way so that states with low Fitness contribute substantially more to the bound on $Q_p$ than high Fitness states (Eq. 4). The denominators in these equations ensure that the values of $F_s$ and $Q_p$ are normalized at each iteration $n$. Eqns. 3 and 4 are iterated over a total of $N$ iterations to obtain the fixed-point values, and for this work we use $N = 100$. Overall, economic complexity provides us a unique, non-linear, non-parametric approach to explore the heterogenous dynamics of economic development.

For the purpose of this analysis, we use state level export data of goods for 12 Indian states (Andhra Pradesh, Bihar, Delhi, Goa, Gujarat, Karnataka, Maharashtra, Odisha, Punjab, Tamil Nadu, Uttar Pradesh, and West Bengal) for which data is available from the Government of India for the period 2009-10 to 2016-17 across a consistent set of 165 products (data purchased from indiastat.com). Together, these states account for over 72% of the country's population. This data does not capture service exports, meaning that some of the high value export sectors such as software and tourism remain outside the ambit of this analysis. While a comprehensive analysis would ideally require data across all states and sectors, the available dataset provides significant population, product, geographic and economic diversity to still remain valuable for analysis.

## 3. Results of Economic Complexity for Indian states:

We start by creating the state-product matrix ($M_{sp}$) based on Revealed Comparative Advantage and find that this describes a triangular matrix (Fig. 1A), implying that the states that have higher Fitness have a more diverse basket of exports – meaning that the set of capabilities they possess enables them to produce all products that fall within that capability limit. The lowest Fitness states on the other hand can only produce a very limited set of products, requiring only a very small set of capabilities. This is in contradiction to Ricardo's thesis of economic specialization (Ricardo, 1817), and in fact suggests that in a dynamic economic environment, states with greater capabilities tend towards greater product diversity (and not increasing specialization), and consequently greater adaptability in the face of varying economic conditions. This outcome is in close agreement with the findings of Hidalgo and Hausmann (2009) as well as Cristelli, Tacchella, Cader, Roster, & Pietronero (2017) who construct country-product matrices using multiple cross-country trade data sets and find the emergence of a triangular country-product structure. Essentially, this result indicates that product basket diversity is attained through the continued enhancement of productive capabilities and that states with low levels of capability might be left competing only in those products where most other states are actively competing as well.

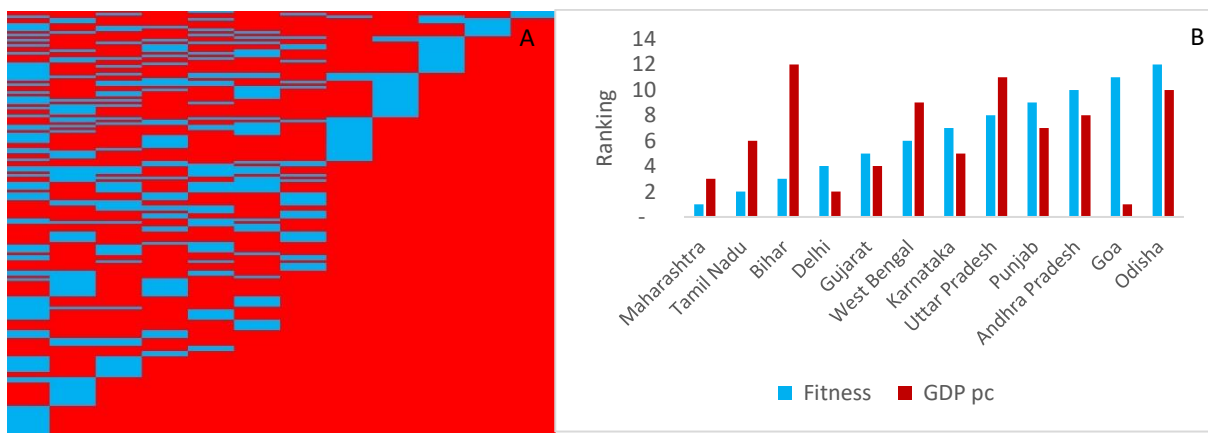

*Figure 1*: **State Product Matrix and Ranking of States (2016-17)**: A: State-Product matrix ($M$), obtained by ordering products by increasing Complexity (on the rows) and ordering states by decreasing Fitness (on the columns), reveals a triangular structure. The blue cells indicate $M_{sp} = 1$ meaning that a state has a relative advantage for a given product, while red cells indicate $M_{sp} = 0$. The triangular structure illustrates that states export most products that fall within their capability set, and the lower the Fitness of a state, the smaller its capability set and consequently its export basket. B: Ranking of States by Fitness and Income (2016-17). There are significant discrepancies in the rankings of states by Fitness (blue) and per capita income (red). This suggests that the economic dynamics revealed by Fitness are different from those revealed by income, though both these variables capture real economic performance.

Fig. 1B compares the ranking of states by their economic complexity (Fitness) and income per capita. Maharashtra is the state with highest Fitness, followed by Tamil Nadu, while Odisha and Goa are the states with lowest Fitness. Some of the striking findings that emerge from these rankings are the high level of Fitness exhibited by Bihar (ranked 3rd), and the much lower levels of Fitness of Punjab (9th) and Andhra Pradesh (10th). We contextualize these findings in the next section. As discussed earlier, Fitness rankings are meant to reveal the extent of capabilities developed by states, with higher Fitness indicating the availability of highly complex capabilities. In order to assess these capabilities of states, we attempt to specify the probability of coexistence of each pair of productive capabilities (manifested in products) at a given time. Hidalgo, Klinger, Barabási, & Hausmann (2007) propose a network of product relatedness termed the 'product space', where relatedness of or similarity between products $i$ and $j$ at time $t$ (similarity in this context would imply requirements of similar underlying infrastructure, institutions, technology and skills) is the conditional probability of both products having $RCA \geq 1$ at time $t$. The product space ($\varphi$) is therefore a square matrix of dimension $PXP$, where $P$ is the total number of products in the export basket and is representative of the complete set of underlying capabilities required to make those products. Each element $\varphi(i,j,t)$ of the product space is given by Eq. 5:

$$\varphi(i,j,t) = \min \{P(RCAx_{i,t} \geq 1 \mid RCAx_{j,t} \geq 1), \ P(RCAx_{j,t} \geq 1 \mid RCAx_{i,t} \geq 1)\} \quad (5)$$

Constructing the product space matrix for our dataset (Fig. 2A), we find that it is a sparse matrix with 15% of its elements equal to 0, 28% less than 0.1, and 45% less than 0.2. These results are consistent with the product space obtained using global trade data (Hidalgo, Klinger, Barabási, & Hausmann, 2007). As Fig. 2A suggests, the probabilities of coexisting capabilities are higher along the diagonal of the matrix, and given that the products are ordered by Complexity, this implies that coexistence is more likely between locations on the product space that have small differences in Complexity. This is brought into even sharper relief in Figs. 2B and 2C, which only highlight those cells in the matrix whose probabilities $\varphi(i,j) \geq 0.5$ and $\varphi(i,j) \geq 0.6$ respectively, and it is readily apparent that higher coexistence probabilities are clustered around the diagonal.

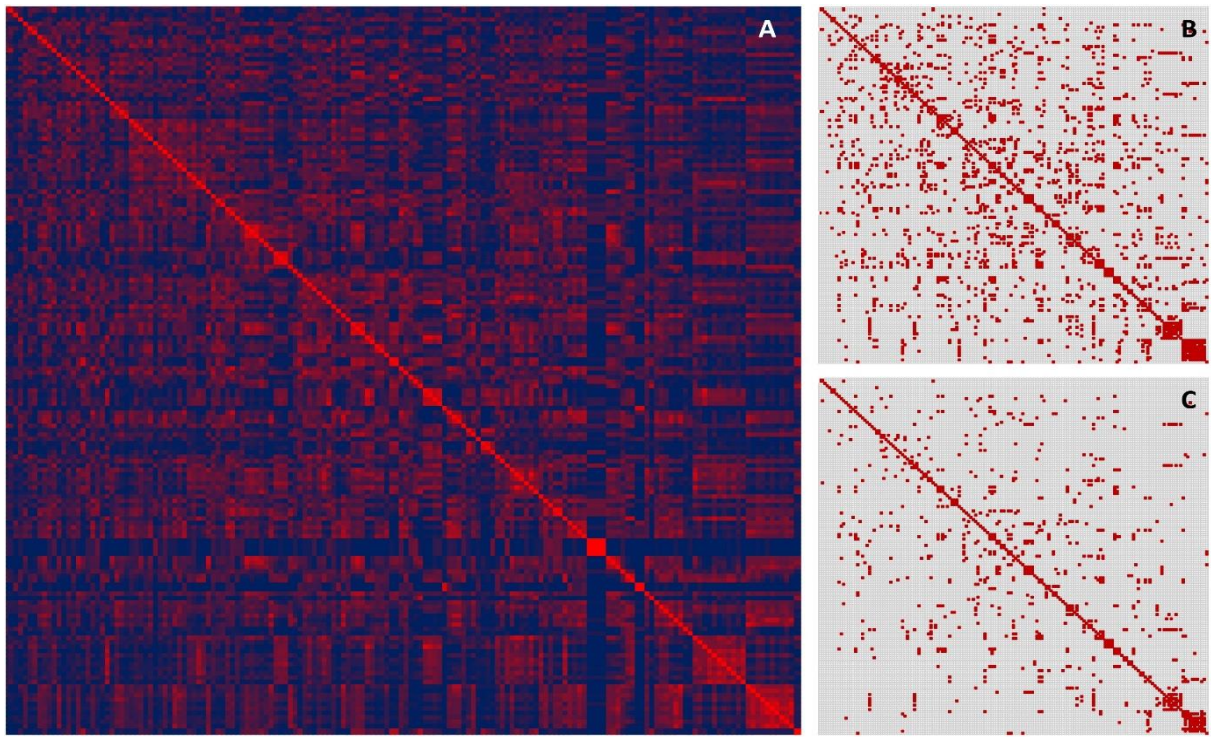

*Figure 2*: **Product Space ($\varphi$)**: A: Product space matrix comprising the probabilities of coexistence of pairs of products. Probability values increase from blue to red. Probabilities for coexistence appear to be maximised around the diagonal of the matrix, indicating the higher likelihood of coexistence between products of similar Complexity. B: Displays in red only those pairs of products with $\varphi(i,j) \geq 0.5$. C: Displays in red only those pairs of products with $\varphi(i,j) \geq 0.6$.

In the context of states will small sets of capabilities, these combinations most likely reflect clusters of co-located, low Complexity capabilities. This essentially points to the possibility of long-term 'low Fitness' traps, where the dynamics of preferential attachment ensure that states with small capability sets remain at sub-optimal levels of Fitness over long periods of time. This is in keeping with our finding in Fig. 3A, where the lowest Fitness states retain their position at the bottom of Fitness rankings from 2009-10 to 2016-17.

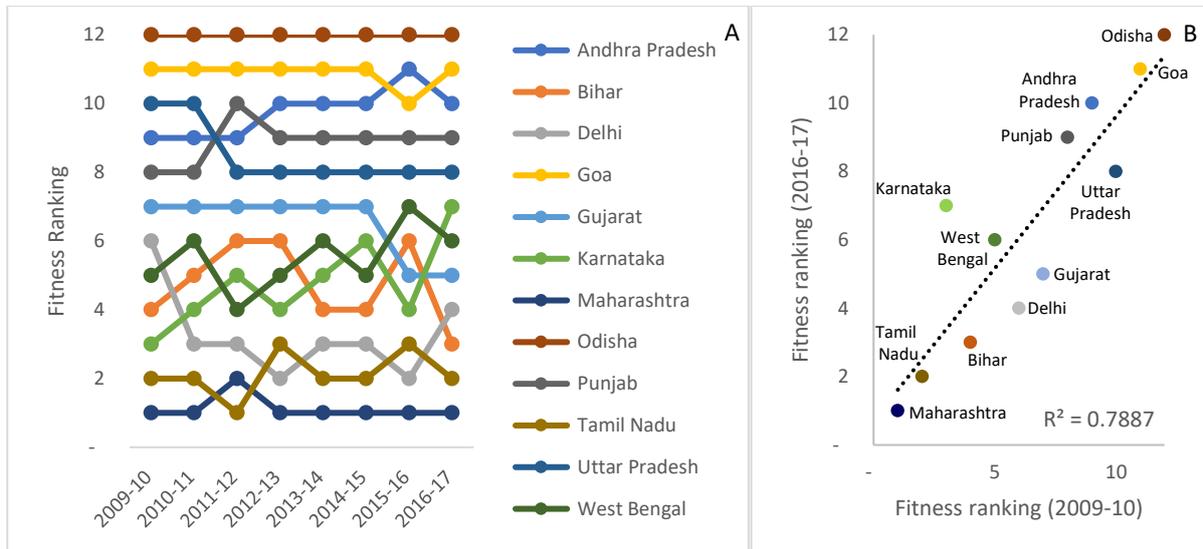

*Figure 3*: **Robustness of Fitness:** A: Temporal evolution of Fitness rankings (2009-10 to 2016-17). The most and least Fit states retain rankings over time, while those in the middle show relatively more Fitness rankings change. C: Fitness Ranking (2009-10) v. Fitness Ranking (2016-17): There is significant correlation between rankings of states over time, with correlation coefficient $r^2 = 0.79$.

We further explore the temporal evolution of Fitness over the period 2009-10 to 2016-17 and find that the most and least Fit states essentially retain their rankings over the 8-year period, while there is more churn in rankings of states in between (Fig. 3A). For instance, Maharashtra and Tamil Nadu are the most Fit states, while Odisha and Goa are the least Fit states, at the beginning and end of the period under consideration. This is potentially a reflection of the fact that the most Fit states have developed highly complex capabilities that are not easily replicated by others, and the lowest fitness states have such a small capability set that any improvement in this set will require long time periods. Fitness rankings in 2009-10 and 2016-17 are highly correlated, with correlation coefficient $r^2 = 0.79$ (Fig. 3B), a result that echoes findings from Chinese regions (Gao & Zhao, 2018). This is a clear indication that the temporal evolution of Fitness of states is relatively stable and gradual, just as we would expect given the progressively greater difficulty in developing more complex capabilities. Overall, this analysis suggests Fitness is a robust measure, representing a real underlying economic dynamic of Indian states.

**4. Contextualizing Fitness using Human Development:**

We now seek to contextualize the emergent Fitness measures of Indian states using a broader framework of socio-economic development. Specifically, we propose to use the Human Development Index (HDI), which is a statistical measure that equally weights education (literacy and school enrolment), health (life expectancy) and standard of living (income) to derive a composite indicator of human development (UNDP, 2019; Stanton, 2007). HDI emphasizes the importance of enhancing capabilities of individuals to live better lives with freedom and opportunity, over a singular focus on the means (income) to achieve these ends (Sen, 1980; Ul-Haq, 1995; Bagolin & Comim, 2008; Adrogué & Crespo, 2019). Indeed, there is evidence to suggest that the relationship between human development and economic development extends in both directions, with investments in education and health from economic growth driving human development, and investment rate and income distribution effects flowing from human development to economic growth (Ranis, Stewart, & Ramirez, 2000). Human development is also found to play an essential role in determining growth trajectories (Suri, Boozer, Ranis, & Stewart, 2011). Empirical work indicates that there are implications of sequencing, with countries that focus on economic growth initially showing a greater likelihood of falling into a vicious cycle of low human development and poor economic growth, while those focused on human development initially, more likely to get on a virtuous cycle of high human development and improved growth prospects (Ranis, Stewart, & Ramirez, 2000). Therefore, successful policy appears predicated on an initial focus on human development, both for its direct impact on individuals and its feedback effect on sustaining economic growth (Ranis, Stewart, & Ramirez, 2000; Suri, Boozer, Ranis, &

Stewart, 2011). Evidence from Indian states also suggests two-way causality between human and economic development (Dholakia, 2003; Ghosh, 2006). HDI, which encapsulates outcomes on education, health, and income, represents the performance of states in ensuring investments that unlock capacities of individuals – and it is arguably these capacities that potentially underlie the development of (some of the) capabilities revealed by economic complexity.

Given this context, we now explore the co-evolution of human development and Fitness of Indian states and study these dynamics for insights into factors impacting economic performance as well as future prospects of individual Indian states. Fig. 4 plots the evolution of Indian states on the HDI - Fitness plane for the period 2009-10 to 2016-17, and we find the emergence of 4 regimes on this plane. As far as Fitness is concerned, we find a laminar regime for $\log(Fitness) > 0$ which reveals predictable co-evolutionary dynamics between Fitness and HDI over time, and a chaotic regime for $\log(Fitness) \leq 0$, where the paths show high variability and offer little scope for predictability. For HDI, we find greater predictability in the region where $HDI \geq 0.68$, which we find is very close to the Human Development Report's definition of high Fitness ($HDI = 0.7$). Therefore, the 4 regimes on the HDI-Fitness plane are: the low HDI - low Fitness quadrant with $HDI \leq 0.68$ and $log(Fitness) \leq 0$ (regime 1), the high HDI- low Fitness quadrant with $HDI > 0.68$ and $log(Fitness) \leq 0$ (regime 2), the high HDI – high Fitness regime with $HDI > 0.68$ and $log(Fitness) > 0$ (regime 3), the low HDI – high Fitness regime with $HDI \leq 0.68$ and $log(Fitness) > 0$ (regime 4).

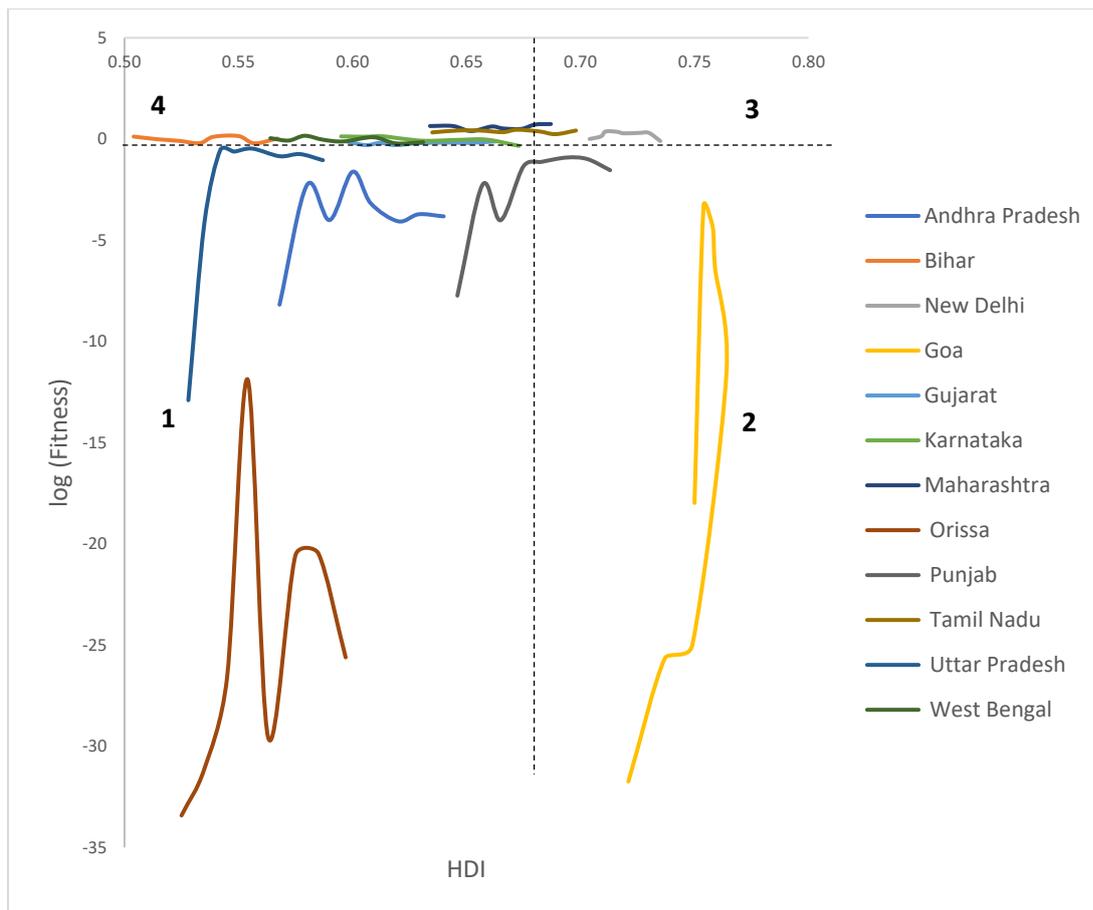

*Figure 4*: **The HDI - Fitness Plane (2009-10 to 2016-17)**: HDI v. log (Fitness). The paths described by 12 Indian states on the HDI - Fitness plane reveal four distinct regimes in the plane. Regime 1 is the low HDI – low Fitness region; regime 2 is the high HDI – low Fitness region; regime 3 is the high HDI – high Fitness region; and regime 4 is the low HDI - high Fitness region.

As of 2016-17, there are three states in the high HDI – high Fitness quadrant (regime 3, Fig. 4) – Delhi, Tamil Nadu, and Maharashtra - and these states are arguably best positioned to benefit from the virtuous cycle discussed earlier between economic development (captured here by Fitness) and human development (HDI). In order to probe this result further, we look at the detailed and complete production spectrums (Pietronero, Cristelli, & Tacchella, 2013) of each of the states (Fig. 5), which essentially captures the volumes of export for each product for each state, ordered by increasing Complexity. We find that Maharashtra and Tamil Nadu have the widest product spectrums followed by Delhi, with Maharashtra and Tamil Nadu producing almost the entire basket of goods, though Maharashtra produces higher volumes than Tamil Nadu. Additionally, while state-level time-series data of service exports is unavailable, it is known that both these states are also significant contributors to export in high value services such as software (van Dijk, 2003; Kambhampati, 2002) - reflecting both their more advanced production capabilities as well and their ability to leverage these capacities to expand into the knowledge economy space for continued economic growth.

There are two states, Goa and Punjab, in the high HDI – low Fitness quadrant (regime 2, Fig. 4), which indicates that their human development should enable them to improve economic performance (Fitness). A deeper examination of these cases however reveals significant differences between them. The production spectrum of Goa is extremely thin and concentrated on the lowest complexity products, explaining its low Fitness (Fig. 5), but it has a significant service economy related to travel and tourism (D'Mello, et al., 2014), which is not captured in the data here. Additionally, it is one of the smallest states in India with a population of 1.46 million in 2011 (Census of India, 2011), which potentially mitigates against the development of a wide production spectrum like Maharashtra or Tamil Nadu.

Punjab, on the other hand, with a population of ~28 million in 2011 (Census of India, 2011) presents a more worrying picture, with its production spectrum essentially a flat line (Fig. 5). The state had a legacy of the remarkable agricultural performance post the Green Revolution of the 1960's, though more recently there have even been worries about stagnation in the state's agrarian economy (Sidhu, 2002). Additionally, it has been pointed out that after the economic reforms of 1991, there has been a deceleration in growth of an already limited industrial economy of the state (Singh, 2006). It is no surprise therefore, that the Fitness of Punjab is indeed quite low. While Punjab's HDI does provide some optimism for the future, it would appear that state's lack of investment in social and economic infrastructure (Singh & Singh, 2002) has ensured that human capacities so developed cannot meaningfully benefit the state.

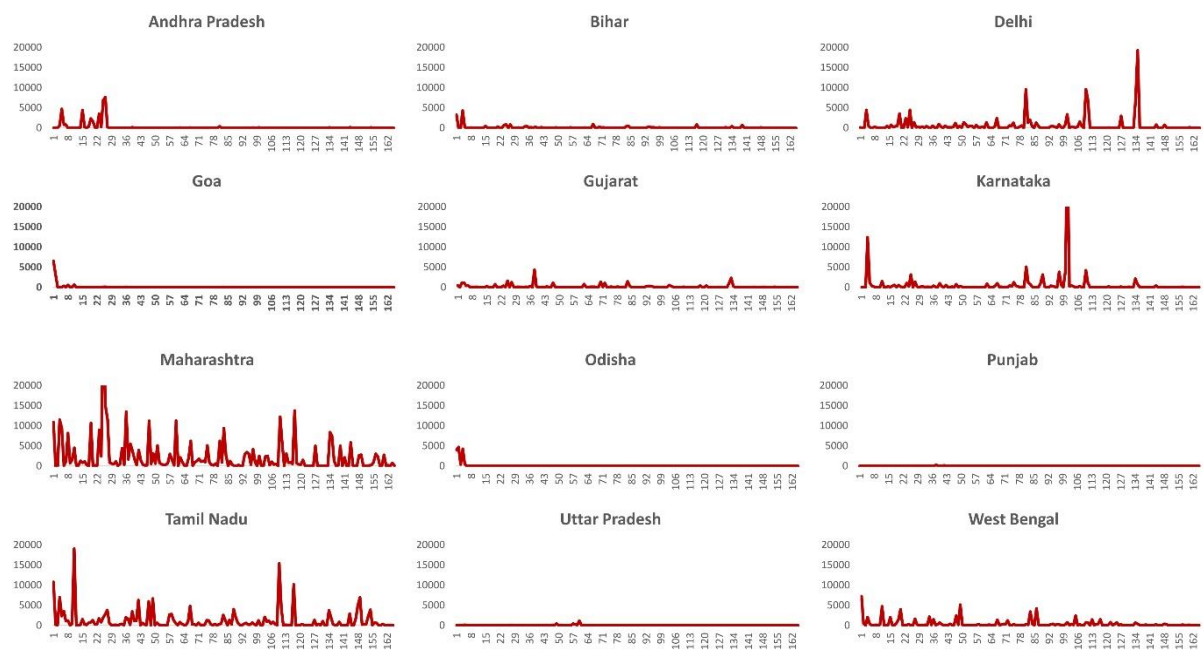

*Figure 5*: **Product Spectroscopy of States (2016-17)**: Products ordered by increasing Complexity v. Volume of export (INR). Shows 12 product spectrums, one for each individual state in the analysis. The wider the spectrum, the greater the Fitness of the state because of its ability to build more Complex products. Maharashtra and Tamil Nadu have the widest spectrums (and highest Fitness), while Odisha, Uttar Pradesh, and Goa have very thin spectrums (indicative of low Fitness).

Next, we consider states in the low HDI – high Fitness quadrant (regime 4, Fig. 4) – West Bengal, Karnataka, Gujarat, and Bihar. Karnataka shows a reasonable spread of capabilities (Fig. 5), similar to Delhi, but has lower Fitness because its production spectrum flatlines earlier as we head towards the highest Complexity products. It is important, however, to remember that the export data does not include services, and Karnataka has been the national leader in export of software services (van Dijk, 2003). To illustrate, while data on software exports disaggregated at the state-level is unavailable post 2007-08, as of that year software exports from Karnataka were 1.4 times the total product export basket of the state. This highlights the development of capabilities in technology and IT which are potential drivers for improved economic performance in high value services in the future and reflect the emergence of productive capabilities at the cutting edge of the knowledge economy. Also, Karnataka's HDI is very close to entering the high region ($HDI \geq 0.7$), which means that with continued improvements in human development, the state is poised to move into the high HDI - high Fitness regime (regime 3, Fig. 4).

West Bengal also has a reasonable spectrum of capabilities (Fig. 5), but for a state of it size, containing 7.6% of India's population (Census of India, 2011), the volume of production across the range of products appears low, when compared to Maharashtra or Tamil Nadu. Thus, even as it is shows comparative advantage in producing several goods, it appears unable to maximize scale economies in the production of these goods. This is reflected in the decline of its manufacturing sector (Lahiri & Yi, 2009), with its national export share in medium technology products dropping to ~8% and high technology dropping to ~4% (Pradhan & Das, 2016). With relatively low HDI (below all-India HDI) and declining export prowess, there is a real risk that the state could slip into the regime of low HDI – low Fitness unless it invests significantly in human development and also an creates enabling environment for industries, large and small, to benefit from citizens with enhanced capabilities.

A similar concern emerges for Gujarat as well, whose production spectrum is narrower than West Bengal, and with low volumes in the products that it exports (Fig. 5). This is quite surprising given Gujarat's position as one of the predominant exporting states in India, accounting for ~15% of India's exports by value (Industries Commissionerate Gujarat, 2020). However, close examination of the export data reveals that while value of exports from Gujarat may be high, they are concentrated in a smaller set of products than Maharashtra, Tamil Nadu, and even West Bengal. Indeed, assessing industrial composition of the state, it was found that chemicals and petroleum constituted close to half the net value added by major industry groups, and that excluding petroleum refining, the contribution of the rest of the manufacturing sector had actually declined (Nagaraj & Pandey, 2013; Pradhan & Das, 2016). This situation has also been exacerbated by the decline of micro, small, and medium enterprises in the state (Pradhan & Das, 2016). Gujarat needs further improvement in its human development to enter the high HDI category, and only continued investment in this direction coupled with diversifying into neighbouring categories in the product space can enable the state to transition into the high HDI – high Fitness regime over time.

Now, we come to the case of Bihar, whose high Fitness is anomalous largely on account of its small overall contribution to the total export basket (2.1%, just slightly ahead of Odisha, Uttar Pradesh, and Goa) – a contribution which is spread across small volumes of a range of products (the blips in Fig. 5). This essentially means that Bihar is able to export a spectrum of products at very small scale, but unable to meaningfully scale up even though it is the third largest Indian state by population (Census of India, 2011). Given that it has the lowest HDI of any Indian state, and that its Fitness is largely a quirk due to the definition of *RCA*, its position is best understood as belonging in the low HDI – low Fitness quadrant. This is a state in need of quite significant investments in human development over time before it can even begin to draw benefits in terms of enhanced economic complexity.

Finally, we consider the states that are of greatest concern, those belonging to the low HDI – low Fitness regime (regime 1), namely Odisha, Uttar Pradesh, and Andhra Pradesh. The presence of Andhra Pradesh (AP) in this quadrant is surprising, but even a cursory examination of its product spectrum (Fig. 5) reveals that AP's exports constitute a set of very low complexity products. AP, while still largely agrarian and with a long history of commercialised agrarian economy (Upadhya, 1988), has a seen a significant shift in economic composition post the emergence of the Information Technology (IT) and software sector in its economy (Dabla, 2004). For instance, in 2007-08, the total computer software and services export from AP was 1.8 times the total product export basket of the state. It has also been argued that as part of this thrust towards IT in AP, the development of human capital in the high-technology and knowledge economy sectors through promotion of technical institutes of higher education over a period of time has been a key policy prerogative of the state government (Dabla, 2004). This analysis suggests that the sustained investments made by AP in improving HDI need to continue into the future for it to sustainably transition into a higher complexity economy. It is important to note that the state of AP was bifurcated into two states, AP and Telengana, in 2014 and we will need longer time series from both these states to meaningfully discuss the evolution of their economic complexity.

The most significant concerns relate to the states of Odisha and Uttar Pradesh where we see almost completely flat product spectrums reflective of abysmal levels of Fitness (Fig. 5). The dramatic fluctuations in Fitness in this chaotic regime (regime 1, Fig. 4) suggest that even small changes in the *RCA* profile of these states yields large Fitness impacts because a single additional product added to or removed from the state's export basket the has a significant impact on an already small state product basket. In this context of low Fitness states, the most critical concern is the ability of these states to develop multiple sets of increasingly complex capabilities, but as we saw earlier while describing the product space (Fig. 3), the probability of coexisting capabilities is maximised when they are closer together in the produce space. Given also the very low levels of HDI in these states, and the need for significant investment to enhance these levels, there is a real risk that these two states (along with Bihar), could be stuck in long-term poverty traps. What this suggests for states stuck in the low HDI - low Fitness regime is that there are no alternatives to long-term strategies focused on building human and physical capital that will enable the systematic creation of increasingly complex sets of capabilities over time.

We are aware that economic complexity potentially reveals a different underlying economic dynamic to that captured by income (Fig. 1B), and we attempt to delineate the nature of these differing dynamics by comparing our results with those of Ghosh (2006), who studied change in income and HDI in Indian states for 1981, 1991, and 2001. There are 10 states that are common to both analyses. We find Maharashtra and Tamil Nadu amongst the best performers and Odisha, Uttar Pradesh, and Andhra Pradesh among the worst performers in both analyses, reflecting the fact that these virtuous and vicious cycles can be hard to break out of, and have long-term repercussions. The discrepancy between dynamics of economic complexity and income however becomes apparent when we compare some of the other states. First, let us consider the cases of Gujarat, which has higher income (also evident from current data, Fig. 1B), and Tamil Nadu, which has higher Fitness. The high average income of Gujarat masks the skewed nature of economic development and limited export basket we discussed earlier (which is however captured by Fitness), while the higher Fitness of Tamil Nadu (and lower income) could encapsulate the unexpressed potential for income growth in the state. Second, we consider Karnataka, which has higher Fitness and income currently, and Punjab, which had higher income in 2001. Punjab's average income is largely a legacy of the Green Revolution and not indicative of its current productive capabilities as revealed by its Fitness measure, while Karnataka's Fitness indicates its production spectrum encompassing even some higher Complexity products, which has enabled it to improve its income levels over time. Finally, we discuss the anomaly of Bihar, which is found to have reasonably high Fitness and low income currently, and had low income in 2001 as well. In this case, it is the Fitness measure that is misleading, as we have discussed before, and income which is more revealing about its economic standing.

## 5. Conclusion:

We attempt to explore the economic complexity of Indian states using goods export data and find that the State-Product matrix, based on product exports in which states have Revealed Comparative Advantage, yields a triangular matrix indicative of the fact that states produce most products for which they have the capability. Thus, states with the ability to produce more Complex products display greater Fitness, and this conception of Fitness encompasses both product diversification and flexibility in a dynamic sense. In this context, we examine the probability matrix of the coexistence of pairs of capabilities and find that the probabilities are maximised when the capabilities are of similar Complexity. We find Fitness to be a robust measure, and reflecting a true underlying economic dynamic.

We then contextualize the economic Fitness measures using Human Development Indicators (HDI) of states. HDI is known to have a two-way causal relationship with economic growth, and an initial focus on improving HDI has been empirically shown to impact economic growth. Exploring the temporal paths described by states on the HDI - Fitness plane reveals four regimes of HDI-Fitness. Maharashtra, Tamil Nadu, and Delhi are in the high HDI- high Fitness quadrant indicative of states on the path to developing a virtuous cycle between human development and economic growth. Goa and Punjab fall into the high HDI – low Fitness regime, with the potential to translate investment in capacities of individuals into economic benefits for the state. Punjab, however, is a concern because of falling investments in social and physical infrastructure, which will be essential to translate human development gains into economic gains. Karnataka, Gujarat, West Bengal, and Bihar fall into the low HDI – high Fitness regime, with the inherent risk of poor development of human capabilities impacting sustained economic development. Amongst these states, Karnataka appears best poised to transition to the high HDI – high Fitness quadrant. Gujarat's export basket is reliant on a smaller set of products and improved economic complexity will require both investments in human development and concerted efforts to expand into adjacencies in the product space. West Bengal's diminishing export profile and lower HDI make it a real risk for transitioning into the low HDI – low Fitness regime. Bihar has the lowest HDI among all Indian states and its Fitness is an anomaly due to the definition of *RCA* – it belongs firmly in the low HDI – low Fitness regime, along with Odisha and UP. It is these three states that are of most concern because their very low HDI and the prospect of being stuck in long-term poverty traps. Long-term planning and investment horizons are required to meaningfully enhance the capabilities of the poorest and least Fit states.

Our work has some important limitations. It considers available data for Indian states, but this is restricted to a time-series of 12 states between 2009-10 and 2016-17. The data set also includes only goods and commodities export, not the export of services, therefore missing out on high value exports such as software. And finally, it only offers an analysis of economic complexity of Indian states, but does not look at how these states fare vis-à-vis the economic complexity of competing nations around the world. These are strands for future work.

Finally, we explore the different underlying economic dynamics that are revealed by economic complexity and income, and discuss the possibility that these discrepancies could encapsulate aspects such as the unexpressed potential for income growth in some cases such as Tamil Nadu, Karnataka, and Maharashtra, and over-reliance on specific markets, products, or historical factors driving income growth in others such as Goa, Gujarat, and Punjab.

**Conflict of Interest:** The authors declare no conflict of interest.
**Competing Interests:** The authors declare that they have no competing interests.
**Funding:** The authors received no funding for this work.